\documentstyle[12pt,aasms4,psfig]{article}
\begin{document}
\title{On the large  escape of ionizing radiation from
GEHRs} \author{Marcelo Castellanos\altaffilmark{1}, {\'A}ngeles I. D{\'\i}az\altaffilmark{2}
\affil{Dpt. F\'{\i}sica Te{\'o}rica C-XI, 
Universidad Aut{\'o}noma de Madrid, \\
E-28049 Cantoblanco, 
Madrid, Spain\altaffiltext{1}{e-mail: marcelo@pollux.ft.uam.es}
\altaffiltext{2}{e-mail: angeles@pollux.ft.uam.es}}
Guillermo Tenorio-Tagle\altaffilmark{3,4}\affil{INAOE,\\ 
Apartado de Correos 51 y 216, 72000 Puebla, M\'exico}
\altaffiltext{3}{e-mail: gtt@inaoep.mx}
\altaffiltext{4}{IBERDROLA Visiting Professor for Science and Technology, U.A.M.}}
\begin{abstract}
A thorough analysis of well studied giant HII regions on galactic discs
for which we know the
ionizing stellar population, the gas metallicity
and the Wolf-Rayet population, leads to photoionization
models which can only match all observed line intensity ratios ([OIII] ,
[OII] , [NII], [SII] and [SIII] with respect to the intensity of H$\beta$), as
well as the H$\beta$ luminosity and equivalent width if one allows for an important escape of energetic ionizing
radiation. For the three regions presented here, the fractions of escaping Lyman continuum
photons amount to  10 to 73 \% and, in all cases, the larger
fraction of escaping photons has energies between 13.6 and 24.4 eV. These escaping photons clearly must have an important impact as a
source of ionization of the diffuse ionized gas (DIG) found surrounding many
galaxies, as well as of the intergalactic medium (IGM).  
\end{abstract} 
 \section{INTRODUCTION}
The issue of whether or not ionizing star clusters are the sources of 
radiation responsible for the diffuse ionized gas (DIG) and, moreover, for
the background radiation at low redshift (Heckman et al 2001), is still
controversial. UV observations of four nearby starburst
galaxies (Leitherer et al. 1995; Hurwitz et al. 1997)
showed  that  more than  10\% of the ionizing photons could escape from these
galaxies. On the other hand, Dove et al. (2000) obtained  from theoretical
arguments  that 7\% of photons produced by OB
associations would have to escape into the DIG to sustain it and could also
be consistent with the estimated flux required  to photoionize the Magellanic
Stream (Weiner \& Williams 1996; Bland-Hawthorn \& Maloney 1999). All these
aspects at low redshift are undoubtedly important to establish whether massive
stars in starburst galaxies can rival  QSO's as sources of the
metagalactic background radiation  at high redshifts (z $>$ 3; see Ricotti \&
Shull 2000). Furthermore, an estimate of the escape fraction at  low redshift
is to provide an upper limit to the expected escape fraction at high redshift,
when the universe was much denser.  Beckman et al. (2000) presented evidence,
based on the analysis of the H$\alpha$ luminosity function of HII region
populations in nearby spiral galaxies, that luminous HII regions may be matter
bounded. This would imply that these regions are an important  source
of  photons to the DIG. 
Here  we report strong evidence that points at  Giant
Extragalactic HII Regions  (GEHRs) to be matter bounded, and thus as an
important source of ionization at large distances from their exciting stars.
Our conclusions are based on evolutionary
synthesis models (Leitherer et al. 1999; hereafter ST99) applied to previously
analyzed GEHRs with observed Wolf-Rayet (WR) features (D{\'\i}az et al. 2000;
Castellanos et al. 2002; hereafter referred to as Paper I). The observed
parameters inside these regions, i.e. the fluxes and equivalent widths of the
WR features, the emission line intensities of [OII]($\lambda$3727),
[OIII]($\lambda$5007), [NII]($\lambda$6584), [SII]($\lambda$6717), and
[SIII]($\lambda$9069) relative to H$\beta$, the observed H$\beta$ luminosity and equivalent width, cannot be fitted
simultaneously unless an important fraction of the ionizing radiation  escapes
from these regions.

\section{MODELS AND RESULTS}
The analyzed GEHRs are regions H13 in NGC 628, CDT3
in NGC 1232 and 74C in NGC 4258. The three of them, particularly 74C
and H13, show spectral  signatures  of  prominent Wolf-Rayet features.
The
presence of these stars allows to constrain the age of the ionizing
clusters to between 3Myr and 6Myr. These values  depend on
the stellar mass loss rates and, consequently, on the stellar metallicity that
ultimately controls  when massive O stars enter the Wolf-Rayet phase (Meynet
1995). This metallicity has been assumed to be similar  to that
found through the analysis of the
emission line spectrum (see Paper I).  Once the age and metallicity
are known, a unique Spectral Energy Distribution (SED) can be
provided by evolutionary synthesis models. 

To estimate the age of the ionizing stellar clusters from the
measured strength and equivalent widths 
of the observed WR features we used the models of Schaerer \& Vacca (1998).
Ages between 4 and 4.5 Myr, 3 and 3.5 Myr and 3.5 and 4 Myr are found for
regions H13, CDT3 and 74C with oxygen abundances of 0.2, 0.4 and 0.5 solar,
respectively.
We have then used the ST99 models to predict the SEDs of the ionizing 
clusters.
These models assume  the same stellar tracks as those used by  Schaerer \&
Vacca (1998) and, at the metallicities here considered, 
represent a good approach to the ionizing spectra (see Stasi{\'n}ska et
al. 2001).





The predicted SEDs were then used as input to the latest version of CLOUDY
(Ferland 1999). The models also require of an electron density (assumed to be constant
throughout the nebula),
the derived gas metallicity (see paper I) and an ionization parameter, U =
Q(H)/(4$\pi$R$^2$n$_e$c); where Q(H) is the total flux of ionizing photons
emitted by the cluster per second, R the distance from the ionizing
source to the illuminated face of the cloud, n$_e$ the electron density and  c the speed of light. In spherical
geometry, the average U is proportional to
(Q(H)n$_e\epsilon^2$)$^{1/3}$ where $\epsilon$ is the gas filling factor.
In the case of ionization bounded regions, Q(H) must equal, in the
absence of dust, the number of ionizing photons derived from H$\alpha$
recombinations. Our models have considered values of the  ionization parameter
between  -4.0 $<$ log U $<$ -2.0, through the span of the WR
phase which, combined with the observed numbers of ionizing
photons (Q(H)) imply R
values of the order of 10$^{\rm 20}$ cm - 10$^{\rm 21}$ cm ($\approx$ 30 - 300 pc). The thickness of
the ionized gas shell is of the order of 10$^{\rm 19}$, less than 10 \% of the total
dimensions of the region, which results in a plane-paralel geometry.

Figure 1  shows the run
of the emission line intensities relative to H$\beta$, and the
H$\beta$ equivalent width and luminosity as a function of the ionized gas
shell thickness. The figure also compares the model predictions in each case
with the observed values (horizontal bars).

Figure 1 clearly shows that a consistent fit to all observed
quantities can only be found if the regions are considered to be matter
bounded, i.e. only if there is an important escape of ionizing radiation from
the nebulae. Models in which all photons are absorbed (most right-hand
points in Figure 1) lead to large departures from all observed values.
However,  if one assumes the nebulae to be  matter
bounded, a satisfactory agreement between all predicted quantities and the 
observations is found at a common distance to the ionizing
cluster for each case. Clearly, for these models we have also demanded for 
consistency with the observed H$\alpha$ luminosity which has been found
through an iterative process. Successful models have hydrogen column
densities of the order of 10$^{\rm 19}$ cm$^{\rm -2}$ while ionization bounded models
reach column densities close to 10$^{\rm 20}$ cm$^{\rm -2}$. 

The fraction of escaping photons depends both
on their energy and the  gas column density.  Table 1 gives the main physical
conditions of the observed regions. Columns 1, 2 and 3 identify the GEHRs and
list  the observed metallicity and resultant age of the best fitting models
for the ionizing clusters. Columns 4 indicates the range in ionization
parameter used, and columns 5-7 indicate both, the absolute and relative values
of the photon flux  escaping  in three different energy bins (13.6 - 24.5 eV,
24.5 - 54.4 eV, and larger than 54.4 eV). The log of the total values of 
photons escaping the nebulae range from 49.80 $<$ log Q$_{esc}$(H) $<$ 51.51,
which imply an escape fraction between 10\% (for region H13) and 73\% (for
region 74C).

\section{DISCUSSION AND CONCLUSIONS}

The observed emission lines of GEHR have been widely used to derive the
properties of the ionizing radiation and link them to those of the
stellar populations generating it, but always under the assumption that
the ionized region is ionization bounded. In the absence of an
independent constraint on the age of this stellar population, such as
the presence of WR stars, successful fits to the emission line spectra
of most GEHR have been found within a narrow range of ages, between 2 and
2.5 Myr (Bresolin et al. 1999; Dopita et al. 2000). In fact, if 
similar methods of analysis were applied to the regions modelled here,
we would obtain a similar result. For these regions, however,  
we have accurate determinations of the gas metallicity and their age,
provided by the analysis of the WR spectral features, and therefore we
can synthesize the corresponding SEDs. 
However, the derived  
SEDs are unable to reproduce the observed emission line spectra 
unless the GEHRs  are density bounded. In such a case,  we have shown here that
a satisfactory fit can be found in the three analyzed cases.  

Our results suggest that the leaking of ionizing photons do depend on
the evolutionary stage of the regions. Regions 74C and CDT3 show
the highest escape fractions, consistent with the prominent observed 
WR features in the former spectrum, and the high excitation
lines ([FeII],, [FeIII]) in the latter one. However, region H13, which also
shows a prominent WR feature, shows a far lower escape
fraction which can be understood from its lower metallicity (0.2 solar 
{\it versus} 0.5 solar in 74C and CDT3), affecting the WR phase
strength.  It is plausible that large photon escaping fractions might be 
expected in those regions that experience a hard WR phase, this one
depending on both the age and metallicity.

An important implication from the escape of ionizing photons concerns
the EW(H$\beta$) - age relation, that does not hold for matter bounded
regions. It is a well known fact that few HII regions show EW(H$\beta$) as 
large as those predicted from evolutionary synthesis models (e.g. Melnick,
Terlevich \& Eggleton, 1985; Mas-Hesse \& Kunth 1998). The leak of Lyman
continuum radiation provides a natural explanation in the correct direction.

A close analysis of the energy distribution of  the escaping radiation
(see Table 1) shows that the largest fraction 
has energies between 13.6 and 24.4 eV (up to 80\% in the case of 74C). 
However, if escaping to incident fractions are considered, helium ionizing
photons are clearly dominant (e.g. up to 84\% of the incident photons
with energies between 24.4 and 54.4 eV escape from region
74C). This would explain why the observed emission line spectrum is so
similar to that produced by a star cluster without WR stars.

Furthermore, the derived large amounts of escaping ionizing
photons indicate that GEHRs and stellar OB associations may significantly
contribute to the ionization of the diffuse gas layers observed above the disks of spiral
galaxies.  Though, to our
knowledge, no measurements of the diffuse interstellar gas (DIG) are
available for NGC 628, NGC 1232 and NGC 4258, recent
studies by Ferguson et al. (1996), Oey \& Kennicutt (1997) and Zurita et
al. (2000) stress the
fact that the integrated escaping flux from disk galaxies can attain up
to 10$^{41}$ erg s$^{-1}$,
enough to account for the diffuse H$\alpha$ flux and fully compatible with our derived values for 
single GEHRs.

In our models the lower energy photons emitted by the central star
cluster are more effectively absorbed by the HII region gas while a large
fraction  of the photons with energy between 1.8 to 4 Ry escape the
nebula.
Therefore, the resulting spectral energy distribution of the escaping
photons is slightly harder than that of the ionizing star cluster. For
the three studied regions the Q(He)/Q(H) ratio of the escaping photons
is: 0.22, 0.20 and 0.19 which should be compared to 0.10, 0.14 and 0.15
for the respective ionizing clusters.

One still unsolved problem for the interpretation of
observations of the DIG above the galactic disk is the apparently low
value of the HeI$\lambda$5876 ${\AA}$ /H$\beta$ (Reynolds \& Tufte 1995) which would
be in contradiction with our derived  Q(He)/Q(H) ratios.  Our studied
regions, however, 
are not representative of galactic HII regions ionized by a single star,
but of high luminosity Giant
Extragalactic HII Regions on the disks of spiral galaxies like, for
example, NGC891. For the DIG in this galaxy the average value of
HeI$\lambda$5876 ${\AA}$ /H$\beta$ is 0.1 (Rand 1997). According to Bresolin, Kennicutt
\& Garnett (1999) the value of this ratio for stellar effective
temperatures between 40000 and 50000 K is highly dependent of
metallicity and reaches the value of 0.1 for a metallicity of about 1/4
solar. In fact, most HII galaxies show values of HeI$\lambda$5876 ${\AA}$ /H$\beta$ around
0.10 (Izotov \& Thuan 1998 and references therein).

In summary, by applying evolutionary synthesis models to well-studied GEHRs with observed
WR features, we find that these regions need to be matter bounded in order to
reconcile model predictions with observations. The amount of escaping ionizing
photons per unit time from these regions range from
49.80 $<$ log Q$_{esc}$(H) $<$ 51.51, which implies an escape
fraction between 10\% and 73\% of the available incident photons. These
fractions seem to increase with both the strength of the WR phase and the
metallicity of the ISM. The implication is thus that the mass of the
ionizing clusters is larger than the value derived in a straight manner from the observed 
H$\alpha$ luminosity. It is also clear that the ISM is highly non-uniform and that the
matter swept up in the shells produced by the mechanical energy deposited by
the WR sources and other massive stars is not sufficient to trap the ionization
fronts. This is also the
result 
obtained from numerical calculations that consider both the mechanical
energy
of massive stellar clusters as well as their ionizing luminosity.
Tenorio-Tagle 
et al. (1999) have shown how once the shell of swept up matter becomes
Rayleigh - Taylor unstable and fragments, as it evolves out of a galaxy
disk into the halo 
(the blowout phenomenon), it then allows not only for the venting of the
hot (wind and SN) matter into the halo, but also for the leakage of a
large 
fraction of the ionizing radiation. The latter soon establishes a giant
conical HII
region in the low density halo. The low densities in the halo lead to a
long recombination time and thus to a large leakage of photons into the
IGM. The 
situation changes later, once the shock is able to sweep enough halo
matter,
enhancing locally the number of recombinations in the expanding shell. 
This leads eventually to the trapping of the ionization front. 
Consecuently, 
depending on the stage of the evolution, a large fraction of the
ionizing flux escapes the
nebulae and is freely available to impact upon the gas at large distances from
the host galaxy plane (Collins \& Rand 2001) and it is also likely to escape the
galaxy and cause an important ionization of the IGM. Clearly, further analysis of this
kind must be done to infer whether or not our derived escape fractions are typical of other GEHRs.

\section*{Acknowledgements}
We thank the anonymous referee for many comments and suggestions which significantly
improved the content of the paper.

G.T.-T. is grateful to an IBERDROLA Visiting Professorship 
to UAM during which part of this work was completed.
This work has been partially supported by DGICYT project AYA-2000-0973.


\clearpage

%


%


\begin{table*}
\setcounter{table}{0}
 \begin{minipage}{140mm}
 \caption{Model results}
 \hspace{0.5cm}
 \scriptsize{
 \begin{tabular}{@{}lccccccc@{}}
\hline
\hline
Region & {Z/Z$_{\odot}$\footnote {Gas metallicity relative to solar}} & {t(Myr)\footnote {Derived age from WR population models}}  & log U       & {log Q$_{esc}$\footnote {Columns 5 to 8: Lower and upper values for the number of escaping ionizing photons in the three energy bins and the total one.\\ The brackets in each column  provide the ratios of escaping to incident radiation in each energy bin and for the total one.}}    & log Q$_{esc}$    & log Q$_{esc}$ & log Q$_{esc}$ \\
       &               &         &             & (13.6 - 24.6 eV) & (24.6 - 54.4 eV) & ($\geq$ 54.4 eV) & total \\
\hline
H13    &  0.2          & 4 - 4.5 & -3.00 $\pm$ 0.05 & 49.68 (8.5\%)    & 49.15 (25\%)     & 43.51 (0.1\%) & 49.80 (10\%)  \\
(NGC 628)&             &         &             & 50.17 (26\%)     & 49.41 (45\%)     & 43.70 (0.3\%) & 50.24 (28\%)  \\
\hline
CDT3   &  0.4          & 3 - 3.5 & -3.35 $\pm$ 0.05 & 51.30 (41\%)     & 50.70 (63\%)     & 45.72 (4\%)   & 51.40 (43\%)  \\
(NGC 1232)&            &         &             & 51.38 (48\%)     & 50.74 (68\%)     & 45.76 (5\%)   & 51.47 (51\%)  \\
\hline
74C    &  0.5          & 3.5 - 4 & -3.25 $\pm$ 0.05 & 51.13 (51\%)     & 50.52 (72\%)     & 47.08 (11\%)  & 51.23 (54\%)  \\
(NGC 4258)&            &         &             & 51.42 (71\%)     & 50.74 (84\%)     & 47.60 (26\%)  & 51.51 (73\%)  \\
\hline
 \end{tabular}
}
 \end{minipage}
\end{table*}

\clearpage

%


%


\begin{figure}
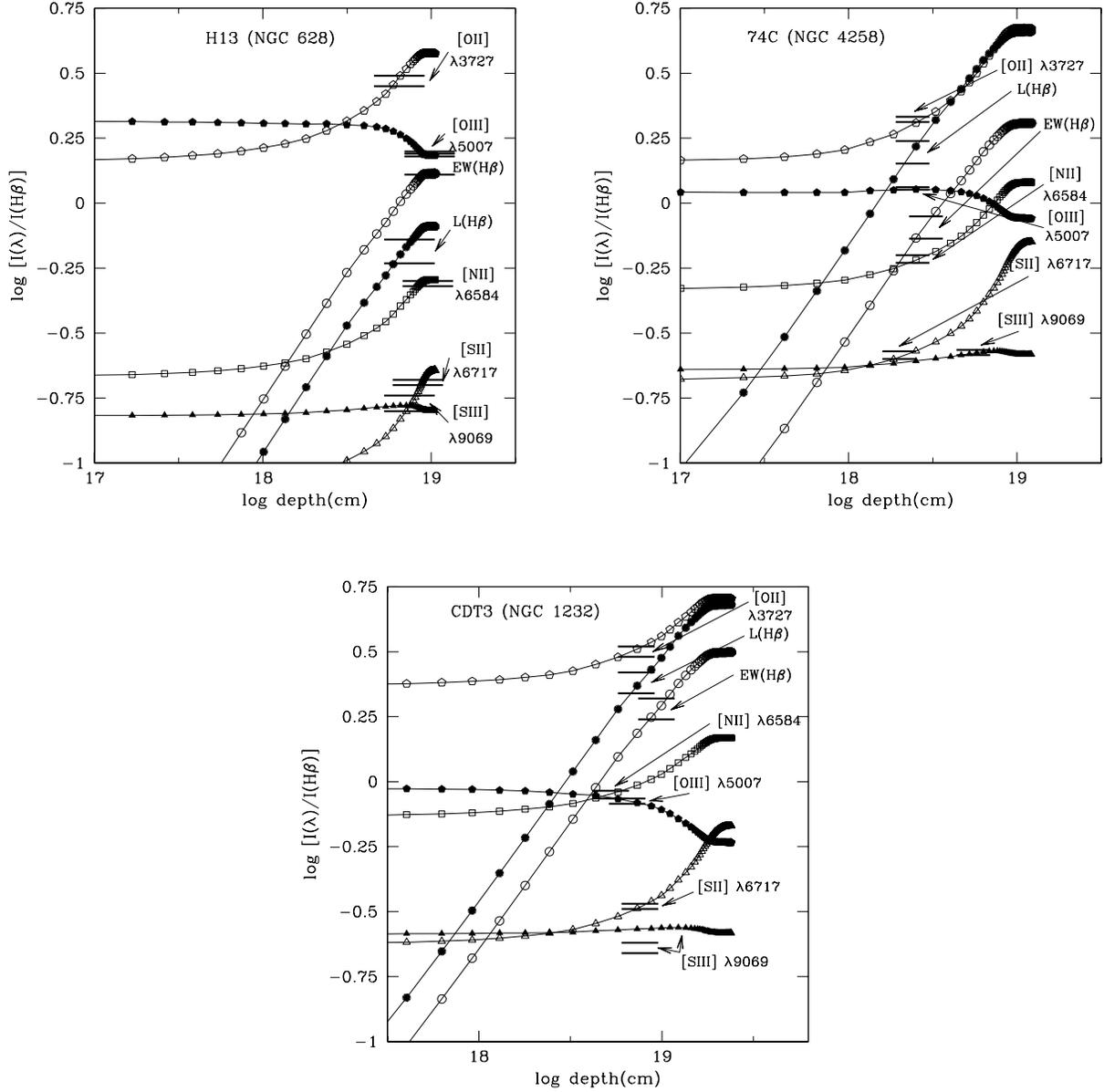

\setcounter{figure}{0}
\centering
\begin{minipage}[c]{20mm}
\centering\psfig{figure=mbh13.epsi,width=7.4cm,height=7.4cm,clip=}
\end{minipage}%
\hspace{1cm}
\begin{minipage}[c]{10mm}
\centering\psfig{figure=mb74c.epsi,width=7.4cm,height=7.4cm,clip=}
\end{minipage}
\\
\vspace{1cm}
\begin{minipage}[c]{10mm}
\centering
\psfig{figure=mbcdt3.epsi,width=7.4cm,height=7.4cm,clip=}
\end{minipage}
\caption{The panels show the run of different line intensity ratios ([OII]($\lambda$3727),
[OIII]($\lambda$5007), [NII]($\lambda$6584), [SII]($\lambda$6717), and
[SIII]($\lambda$9069) relative to H$\beta$), as well as the H$\beta$ equivalent
width and H$\beta$ luminosity (both of them scaled by a constant factor), as
a function of ionized gas shell thickness in the three nebulae here
considered. The models have been derived from the ionization parameter range
given in Table 1 and extend until all photons are used up. The horizontal bars
across every predicted quantity trend indicate the observed values and their
errors (see Paper I).} 
\end{figure}

\end{document}